# Standard Quantum Teleportation of an Arbitrary N-Qubit State, Non-Existence of Magic Basis and Existence of Magic Partial Bases for 2N Entangled Qubit States with N>1

Hari Prakash[1,2*] and Vikram Verma[1**]

[1]Physics Department, University of Allahabad, Allahabad-211002, India
[2]Indian Institute of Information Technology, Allahabad-211012, India

[*]*e-mail: prakash_hari123@rediffmail.com*
[**]*e-mail: vikramverma18@gmail.com*

## Abstract

We present a simple and precise protocol for standard quantum teleportation of N-qubit state, considering the most general resource q-channel and Bell states. We find condition on these states for perfect teleportation and give explicitly the unitary transformation required to be done by Bob for achieving perfect teleportation. We discuss connection of our simple theory with the complicated related work on this subject and with character matrix, transformation, judgment and kernel operators defined in this context. We also prove that the magic basis discussed by Hill and Wootters [Phys. Rev. Lett. **78** (1997) 5022] does not exist for entangled 2N-qubit states with N > 1 but magic partial bases, similar to those discussed recently by Prakash and Maurya [Optics Commun. **284** (2011) 5024] do exist. We give explicitly all magic partial bases for N = 2.

## 1. Introduction

Quantum teleportation (QT) means transfer of information encrypted as q-state of some system with a sender, say, Alice, to a distant receiver, say, Bob without sending the system or any part of information directly. The information is transferred and a similar system with Bob becomes a replica of Alice's system by acquiring the q-state representing the information. Bennett *et al* [1] gave the first protocol for QT of one qubit of information using a quantum channel between the parties, involving sharing of an EPR entangled pair [2] of qubits and a classical 2 c-bit channel for communication by Alice to Bob of result of a Bell state measurement (BSM) by Alice on her two qubits, the one having information encrypted and the one shared by Alice out of the entangled pair. Bob performs a unitary transformation on the q-state of his particle, dependent on the result of BSM, and generates replica of the original q-state on his particle. Quality of QT is decided by fidelity F given by $F = |\langle I | T \rangle|^2$ or $F = Tr[\rho_I \rho_T]$, where $|I\rangle$ and $|T\rangle$ are information and teleported states or $\rho_I$ and $\rho_T$ are the corresponding density operators.

For quantum teleportation (QT) of a single qubit Hill and Wooters [3] noted that if a basis, which they called as the magic basis, with states,

$$|e_0\rangle = (1/\sqrt{2})(|00\rangle + |11\rangle),\ |e_1\rangle = i(1/\sqrt{2})(|00\rangle - |11\rangle),$$
$$|e_2\rangle = i(1/\sqrt{2})(|01\rangle + |10\rangle),\ |e_3\rangle = (1/\sqrt{2})(|01\rangle - |10\rangle)\ ,$$

is defined, and the entangled resource state $|E\rangle$ is expanded as $|E\rangle = \sum_{i=0}^{3} c_i |e_i\rangle$, a parameter *C*, called concurrence, can be written as $C = \left|\sum_{i=0}^{3} c_i^2\right|$ and if $C = 1$, it leads to SQT with $F = 1$. Existence of such magic bases has been reported, in addition to Hill and Wootters [3], only by Prakash and Maurya [4] recently for entangled 3 qubit state in SQT using BSM with 3 entangled qubit states and in CQT using BSM with 2 entangled qubit state when the destinations of the 3-entangled qubits are fixed. These authors note that in other cases similar sets of magic bases with 4 or 2 basis states are obtained and they call these magic semi- bases or magic quarter bases.

QT of information encoded on superposed coherent state has also been studied [5-7] as superposed coherent states are more robust against decoherence [6]. For these studies [5, 7] and for QT using non-maximally entangled states of qubits, it is seen that the fidelity *F* depends on the information state $|I\rangle$ and one has to define the minimum assured fidelity (MASFI) [7], as the minimum value of *F* over the various possible states $|I\rangle$. It can be shown that for QT with superposed coherent states, concurrence $C = 0$ leads to MASFI = 0 and for QT with non-maximally entangled state of 2 qubit MASFI = $2C/(1+C)$ [8].

QT has been realized experimentally [9-11] and also generalized for QT of N qubits [12-16]. It has been shown that for QT of N qubits, the resource has to be entangled state of at least 2N qubits [17]. If entangled 2N qubits are used QT is called standard quantum teleportation (SQT) and if the number of entangled qubits greater than 2N are used and the extra qubits are sent to additional parties, the process is called controlled quantum teleportation (CQT) and it increases the security. Secure exchange of quantum information has been studied recently by Mishra, Maurya, and Prakash [18]

Yang and Guo [19] were the first to study SQT of multi qubits using 4-qubits entangled state. Lee *et al* [20] also studied the same problem. Rigolin [21] studied this problem in great detail and gave a set of 16 generalized Bell states of 4-entangled qubits. Rigolin also described a magic basis but this was different from the Hill-Wootters magic basis in that it does not satisfy the property $\left|\sum_{i=0}^{3} c_i^2\right| = 1$ giving SQT with $F$=1. Rigolin's multi-particle states were shown by Deng [22] to be tensor product of ordinary Bell states. Deng [22] also showed that Rigolin's protocol is in principle the same as the protocol of Yang and Guo [19]. Yeo and Chua [23] gave a protocol for perfect QT of an arbitrary 2-qubit state using genuine multipartite 4-qubit entangled state, which cannot be reduced to a tensor product of two ordinary Bell states. Li *et al* [24] went a step further and gave a protocol for QT of 3-qubit state using genuinely entangled six-qubit state. Zha [25] included involvement of m supervisor also in CQT of 3-qubit.

Zha and Song [26] studied in detail faithful SQT of 2 qubits using 4-qubit entangled state. Their measurement basis are the same as those of Rigolin and they wrote the composite state of six qubits, 2 in information state and 4 in quantum channel, in the form,

$$|Composite\ state\rangle_{123456} = |Information\rangle_{12} \otimes |QuantumChannel\rangle_{3456}$$
$$= \sum_{i,j=0}^{15} |Bell\,states;ij\rangle_{1234}\ \sigma^{r}_{56} |Information\rangle_{56}\ ,$$

and defined $\hat{\sigma}^{ij}_{56}$ as transformation operator(TO). The TO is obviously different for different results of BSM. The authors showed that (i) if TO is unitary QT is perfect but (ii) if TO is not unitary but invertible Bob may use an auxiliary qubit in state $|0\rangle$ as ancilla making the required transformation on his 3 particles. There will be QT with success only if the ancilla is found in the state $|0\rangle$ but failure otherwise, giving success less than unity. The authors also considered non-Bell pair quantum channel. As an example, the authors considered an entangled state which was not factorizable in two Bell states (like Rigolin's g-states) and evaluated the TO. Zha and Ren [27] extended their work further and analyzed the relationship between determinant of TO and a stochastic local operations assisted by classical communication (SLOCC) transformation invariant L and conclude that QT will fail if L is zero. Chen *et al* [28] commented that Zha-Ren protocol [27] is equivalent to Rigolin's protocol [21] in principle, and TO can be used as a means to transform an arbitrary four qubits entangled state into a tensor product of two Bell states. In reply to the comments of Chen et al [28] on their protocol, Zha & Ren [29] remarked that their protocol can be generalized to multipartite and non-symmetric quantum channels and the Rigolin's protocol [21] is only a special case of their protocol. Li *et al* [30] gave a protocol for teleporting an arbitrary three qubits state by using genuine six qubits entangled state. By utilizing the method of Zha and Ren [27], Zhang *et al* [31] have worked out the TO for the case of QT of 3-qubits using an arbitrary six qubits state as quantum channel.

The QT of an arbitrary N qubits state has been studied by many authors [12-16]. Chen et al. [12] gave a protocol for QT of an arbitrary N-qubit state using 2N-qubit entanglement channel which is a tensor product of N-Bell states. Man *et al* [13] have considered the CQT of an arbitrary N-qubit state using (2N+1)-qubit entangled state. Quan *et al* [14] defined a character matrix for a 2N-qubit state and showed that there exist a maximal entanglement between two subsystems of particles 12…N and (N+1) (N+2)……2N if and only if the character matrix is unitary. The character matrix is characteristic of the quantum channel. Ming *et al* [15] gave a criterion for the quantum teleportation of an arbitrary N-particle state using 2N-particle entangled state by introducing a "judgment operator", writing

$$\left|Composite\, state\right\rangle_{12...3N} = \left|Information\right\rangle_{12...N} \otimes \left|Quantum\, channel\right\rangle_{N+1,N+2,.....,3N}$$

$$= \sum_{i_{N+1},i_{N+2},...,i_{2N}=0}^{3} \{\left|MES; i_{N+1} i_{N+2} ... i_{2N}\right\rangle_{123....2N} J^{i_{N+1} i_{N+2} ... i_{2N}}_{2N+1,2N+2,....3N} \left|Information\right\rangle_{2N+1,2N+2,...3N}\},$$

where $J$ is the judgment operator and MES stand for "maximally entangled states". Obviously this judgment operator $J$ is a straight forward genaralization of Zha and Song [26] transformation operator. The $2^{2N}$ Bell states have been formed by application of products of Pauli operators on N-qubits of an entangled 2N-qubit state which is one in the family of Bell states. When the result of BSM is this Bell state, Ming *et al* [15] call their judgment operator as "head judgment operator". Qin *et al* [16] introduced a "Kernel Operator" in teleporting an arbitrary N-qubit state by using 2N-qubit entangled state. A critical examination reveals that this kernel operator is same as the head judgment operator.

In this paper we present a simple and precise protocol for SQT of an arbitrary N-qubit state and prove that the magic basis does not exist for 2N entangled qubits with N > 1 but magic partial bases do exist. In section 2, we present our protocol for SQT and analyse using general quantum channel and measurement bases. We find condition on resource q-channel and Bell states for achieving perfect QT. We also find that the unitary transformation required to be done by Bob for perfect QT and discuss the connection with character matrix [14] transformation operator [26], judgment operator [15] and kernel operator [16]. In section 3, we prove that the Hill and Wootters type magic basis cannot exist for entangled 2N qubit state for N > 1. In section 4, we show further that some magic partial bases do exist for entangled 2N qubit states and give explicitly these for N = 2.

## 2. SQT of an Arbitrary N-Qubit Information State

Let the N-qubit information required to be teleported be encrypted in the state

$$\left|I\right\rangle_{\{I\}} = \sum_{i=0}^{2^N-1} I_i \left|\tilde{i}\right\rangle_{\{I\}}, \tag{2.1}$$

of N qubits $\{I\} = (I_1, I_2, .......I_N)$. Here, for decimal number $i$ which can be written as $i_1 i_2 i_3 .......i_N$ in the binary basis, *i.e.*, $i = \sum_{r=1}^{N} 2^{N-r} i_r$ , the state $|\tilde{i}\rangle_{\{I\}}$ can be written as

$$|\tilde{i}\rangle_{\{I\}} = \prod_{r=1}^{N} |i_r\rangle_{I_r} \quad . \tag{2.2}$$

Coefficients $I_i$ define a $2^{2N}\times 1$ column matrix $\underline{I}$ by

$$(\underline{I})_i = I_i \quad . \tag{2.3}$$

The entangled state $|E\rangle_{\{A\}\{B\}}$ of two subsystems of N qubits, $\{A\} = (A_1 A_2 ......A_N)$ with Alice and $\{B\} = (B_1 B_2 ......B_N)$ with Bob, can be written similarly in the form,

$$|E\rangle_{\{A\}\{B\}} = \sum_{j,k=0}^{2^N-1} E_{jk} |\tilde{j}\rangle_{\{A\}} |\tilde{k}\rangle_{\{B\}} \quad . \tag{2.4}$$

Coefficients $E_{jk}$ help us define a $2^N \times 2^N$ matrix $\underline{E}$ and an operator $\hat{E}$ by

$$(\underline{E})_{jk} = E_{jk} \, , \; \hat{E} = \sum_{j,k} E_{jk} |\tilde{j}\rangle\langle\tilde{k}| \; . \tag{2.5}$$

It may be noted that the matrix $\underline{E}$ is $2^{-N/2}$ times the character matrix defined by Quan *et al* [14].

In the most general case, the $2^{2N}$ Bell states can be written as

$$|B^{(\alpha)}\rangle_{\{I\}\{A\}} = \sum_{i,j=0}^{2^N-1} B_{ij}^{(\alpha)} |\tilde{i}\rangle_{\{I\}} |\tilde{j}\rangle_{\{A\}} \, , \; \alpha = 0, 1, 2, ...., 2^{2N}-1. \tag{2.6}$$

This define matrices $\underline{B}^{(\alpha)}$ and operators $\hat{B}^{(\alpha)}$ by

$$\left(\underline{B}^{(\alpha)}\right)_{ij} = B_{ij}^{(\alpha)} \; , \; \hat{B}^{(\alpha)} = \sum B_{ij}^{(\alpha)} |\tilde{i}\rangle\langle\tilde{j}| \, . \tag{2.7}$$

Normalization conditions give

$$\langle I|I\rangle = \underline{I}^{\dagger}\underline{I} = \underline{1} \; , \; \langle E|E\rangle = Tr(\underline{E}^{\dagger}\underline{E}) = \underline{1} ; \; \langle B^{(\alpha)}|B^{(\alpha)}\rangle = Tr(\underline{B}^{(\alpha)\dagger}\underline{B}^{(\alpha)}) = 1. \tag{2.8}$$

Completeness relation for $|B^{(\alpha)}\rangle$ gives

$$\sum_{\alpha=0}^{2^{2N}-1} |B^{(\alpha)}\rangle\langle B^{(\alpha)}| = \sum_{\alpha=0}^{2^{2N}-1} \sum_{i,j,k,l=0}^{2^N-1} B_{ij}^{(\alpha)} B_{kl}^{(\alpha)*} \left( |\tilde{i}\rangle_{\{I\}}\langle\tilde{k}| \otimes |\tilde{j}\rangle_{\{A\}}\langle\tilde{l}| \right). \tag{2.9}$$

On comparison with resolution of unit operator in the form

$$1 = \sum_{i,j=0}^{2^N-1} |\tilde{i}\rangle_{\{I\}}\langle\tilde{i}| \otimes |\tilde{j}\rangle_{\{A\}}\langle\tilde{j}| = \sum_{i,j,k,l=0}^{2^N-1} \delta_{ik}\delta_{jl} |\tilde{i}\rangle_{\{I\}}\langle\tilde{k}| \otimes |\tilde{j}\rangle_{\{A\}}\langle\tilde{l}| , \tag{2.10}$$

we get

$$\sum_{\alpha=1}^{2^{2N}} B_{ij}^{(\alpha)} B_{kl}^{(\alpha)*} = \delta_{ik}\delta_{jl} \ . \tag{2.11}$$

Composite state of $\{I\}$ $\{A\}$ $\{B\}$ can then be written as

$$|\psi\rangle_{\{I\}\{A\}\{B\}} = |I\rangle_{\{I\}}|E\rangle_{\{A\}\{B\}} = \sum_{i,j,k=0}^{2^N-1} I_i E_{jk} |\tilde{i}\rangle_{\{I\}} |\tilde{j}\rangle_{\{A\}} |\tilde{k}\rangle_{\{B\}}$$

$$= \sum_{\alpha=1}^{2^{2N}} |B^{(\alpha)}\rangle_{\{I\}\{A\}} \sum_{i,j,k=0}^{2^N-1} B_{ij}^{(\alpha)*} I_i E_{jk} |\tilde{k}\rangle_{\{B\}} . \tag{2.12}$$

If the result of BSM is $\alpha$, the Bob's state is

$$|Bob^{(\alpha)}\rangle = C_\alpha \sum_{i,j,k=0}^{2^N-1} B_{ij}^{(\alpha)*} I_i E_{jk} |\tilde{k}\rangle, \tag{2.13}$$

where $C_\alpha$ appears for the sake of normalization of $|Bob^{(\alpha)}\rangle$, and if Bob performs unitary operation $\hat{U}^{(\alpha)}$ given by

$$\hat{U}^{(\alpha)} = \sum_{m,n=0}^{2^N-1} U_{mn}^{(\alpha)} |\tilde{m}\rangle\langle\tilde{n}|, \tag{2.14}$$

which defines matrices $\underline{U}^{(\alpha)}$ by $\left(\underline{U}^{(\alpha)}\right)_{mn} = U_{mn}^{(\alpha)}$, the teleported state is

$$|T^{(\alpha)}\rangle_{\{B\}} = C_\alpha \sum_{i,j,k,m=0}^{2^N-1} B_{ij}^{(\alpha)*} I_i E_{jk} U_{mk} |\tilde{m}\rangle_{\{B\}} . \tag{2.15}$$

If perfect teleportation is possible, $|T^{(\alpha)}\rangle_{\{B\}}$ would be same as $|I\rangle_{\{B\}} = \sum_{m=0}^{2^N-1} I_m |\tilde{m}\rangle_{\{B\}}$ only if

$$C_\alpha \sum_{i,j,k=0}^{2^N-1} B_{ij}^{(\alpha)*} I_i E_{jk} U_{mk}^{(\alpha)} = I_m , \tag{2.16}$$

for arbitrary coefficients $I_i$ . This gives

$$C_\alpha \sum_{j,k=0}^{2^N-1} B_{ij}^{(\alpha)*} E_{jk} U_{mk}^{(\alpha)} = \delta_{im} , \tag{2.17}$$

or, in the matrix notation,

$$C_\alpha \underline{U}^{(\alpha)} \underline{E}^T \underline{B}^{(\alpha)\dagger} = \underline{1}, \tag{2.18}$$

where superscript $T$ denotes transpose. This equation gives

$$\underline{B}^{(\alpha)} = \frac{1}{C_\alpha} \underline{U}^{(\alpha)} \underline{F}^\dagger, \ \underline{F} \equiv \left(\underline{E}^T\right)^{-1}. \tag{2.19}$$

Substitution in Eq.(2.11) then gives

$$\sum_{\alpha=0}^{2^{2N}-1} \frac{1}{\left|C_\alpha\right|^2} \sum_{m,n=0}^{2^N-1} U_{im}^{(\alpha)} F_{mj} U_{kn}^{(\alpha)*} F_{nl}^{*} = \delta_{ik}\delta_{jl} , \tag{2.20}$$

If we put $k$=$i$ and sum over $i$, we get

$$\sum_{\alpha=0}^{2^{2N}-1} \left|C_\alpha\right|^{-2} \left(\underline{F}^{\dagger}\underline{F}\right)_{lj} = 2^N \delta_{jl} \tag{2.21}$$

or $$\underline{F}^{\dagger}\underline{F} = k\underline{1} = \underline{F}\,\underline{F}^{\dagger}, \quad k = 2^N \Big/ \sum_{\alpha=1}^{2^{2N}} \left|C_\alpha\right|^{-2} . \tag{2.22}$$

Using Eqs.(2.19) and (2.22), we get

$$\underline{B}^{(\alpha)}\underline{B}^{(\alpha)\dagger} = k\left|C_\alpha\right|^{-2}\underline{1} = \underline{B}^{(\alpha)\dagger}\underline{B}^{(\alpha)}. \tag{2.23}$$

Using normalization condition $Tr(\underline{E}^{\dagger}\underline{E}) = Tr(\underline{F}^{T}\underline{F}^{*}) = k^{-1}Tr(\underline{1}) = k^{-1}2^N = 1$ and $Tr(\underline{B}^{(\alpha)\dagger}B^{(\alpha)}) = \underline{1}$, we get $\left|C_\alpha\right| = 2^N$ and the conditions

$$\underline{E}^{\dagger}\underline{E} = 2^{-N}\underline{1} \ , \ \underline{B}^{(\alpha)\dagger}\underline{B}^{(\alpha)} = 2^{-N}\underline{1}. \tag{2.24}$$

If we take $C_\alpha$ real and equal to $2^N$, we can write

$$\underline{U}^{(\alpha)} = 2^N \underline{B}^{(\alpha)}\underline{E}^{*}. \tag{2.25}$$

It may be noted that $\hat{U}^{(\alpha)}$ is same as the transformation operator used by Zha and co-workers [27] or the judgment operator used by Ming *et al* [16]. If the Bell states $\left|B^{(\alpha)}\right\rangle$ are derived from some particular Bell state, say, $\left|B^{(0)}\right\rangle$ by application of product of Pauli matrices [15-16] in the form,

$$\left|B^{(\alpha)}\right\rangle = \sigma^{(\alpha)}\left|B^{(0)}\right\rangle, \tag{2.26}$$

where

$$\sigma^{(\alpha)} = \prod_{r=1}^{N} \sigma_{I_r}^{(\alpha_r)} . \tag{2.27}$$

$\alpha_r (= 0,1,2,3)$ give quarternary basis of decimal number $\alpha$, *i.e*, $\alpha = \sum_{r=0}^{N-1} 4^{N-r-1}\alpha_r$ and $\sigma_{I_r}^{(\alpha_r)}$ are $I, \sigma_z, \sigma_x, \sigma_z\sigma_x$ operating on qubit $I_r$ for $\alpha_r = 0,1,2,3$ respectively.

It may be noted that if $\left|B^{(\alpha)}\right\rangle$ have been obtained from $\left|B^{(0)}\right\rangle$ , $\hat{U}^{(0)}$ is head judgment operator of Ming *et al* [15] or the kernel operator of Qin *et al* [16] or the transformation operator $\hat{\sigma}_{56}^{00}$ of Zha and co-workers [27] for N = 2.

**3. Non-Existence of Hill-Wootters Type Magic Bases for Entangled 2N Qubit State with N>1**

Hill and Wootters [3] showed existence of a magic basis for 2 qubit entangled states being used as quantum resource in SQT of single qubit. If we write their magic basis states as $\left|e^{(\alpha)}\right\rangle$ , $(\alpha = 0,1,2,3)$ and write expansion of entangled state in terms of $\left|e^{(\alpha)}\right\rangle$ as $\left|E\right\rangle = \sum_{\alpha=0}^{3} c_\alpha \left|e^{(\alpha)}\right\rangle$, for concurrence $C \equiv \left|\sum_{\alpha=0}^{3} c_\alpha^2\right| = 1$, perfect SQT is obtained with fidelity $F = 1$. In this section, we prove non-existence of magic basis for 2N entangled qubit state for N > 1 by assuming existence of such a magic basis and showing that this leads to inconsistency and absurdness.

If a magic basis with magic basis states $\left|e^{(\alpha)}\right\rangle$ $(\alpha = 0,1,2,...,2^{2N}-1)$ exists, the states define matrices $\underline{e}^{(\alpha)}$ with the property $\underline{e}^{(\alpha)\dagger}\underline{e}^{(\alpha)} = 2^{-N}\underline{1}$ as each state $\left|e^{(\alpha)}\right\rangle$ gives perfect SQT. If the entangled state $\left|E\right\rangle = \sum_{\alpha=0}^{2^{2N}-1} c_\alpha \left|e^{(\alpha)}\right\rangle$ with $c_\alpha$ real and $\sum_{\alpha=0}^{2^{2N}-1} c_\alpha^2 = 1$, $\left|E\right\rangle$ would give perfect SQT and hence $\underline{E}^\dagger \underline{E} = 2^{-N}\underline{1}$.

To make things easy, for $2^N \times 2^N$ matrices, we consider a basis with basis elements $\underline{m}^{(\alpha)}$ for $2^N \times 2^N$ matrices given by direct product of N matrices $\underline{\sigma}_j^{(\alpha_j)} = \underline{I}_j$, $\underline{\sigma}_{jz}$, $\underline{\sigma}_{jz}\,\underline{\sigma}_{jx}$, $\underline{\sigma}_{jx}$ corresponding to $j^{\text{th}}$ qubit, for $\alpha_j = 0,1,2,3$. Thus the $\alpha^{\text{th}}$ element ($\alpha$ = 0, 1, 2, 3,....., $2^{2N}$-1) is $\underline{m}^{(\alpha)} = \prod_{j=1}^{N} \underline{\sigma}_j^{(\alpha_j)}$ where $\alpha_1, \alpha_2, \alpha_3, ......., \alpha_N$ is the decimal number α expressed in quaternary basis, *i.e.*, $\alpha = \sum_{j=1}^{N} \alpha_j 4^{N-j}$. It can be shown very easily* that the elements $\underline{m}^{(\alpha)}$ has the properties that (1) square of each elements is $\underline{1}$ (*i.e.*, $\left(\underline{m}^{(\alpha)}\right)^2 = \underline{1}$), (2) each element is a hermitian matrix (*i.e*,. $\underline{m}^{(\alpha)\dagger} = \underline{m}^{(\alpha)}$), (3) the product of any two elements is $\pm 1$ or $\pm i$ times some other element, (4) any two elements either commute or anti-commute, (5) for any given element $\underline{m}^{(\alpha)}$, we can find at least one element $\underline{m}^{(\beta)}$ such that $\underline{m}^{(\alpha)}$ and $\underline{m}^{(\beta)}$ anticommute, (6) trace of element $\underline{m}^{(\alpha)}$ is zero for $\alpha \neq 0$, (7) all matrices $\underline{m}^{(\alpha)}$ are linearly independent and (8) any $2^N \times 2^N$ matrix $\underline{M}$ can be expanded linearly in terms of matrices $\underline{m}^{(\alpha)}$.

Property (4) tells that any two elements either commute or anti-commute but it can be shown that, for N > 1, the situation that all elements $\underline{m}^{(\alpha)}$ with $\alpha \neq 0$ anticommute is impossible. This is clear from the fact that, if we consider the product $\underline{m}^{(1)}\underline{m}^{(2)}$, which is a member of $\underline{m}^{(\alpha)}$ family (property 3), it does not commute with $\underline{m}^{(3)}, \underline{m}^{(4)}, \ldots\ldots, \underline{m}^{(2^{2N}-1)}$. For N = 1, it so turns out that $\underline{m}^{(0)} = \underline{1}$, $\underline{m}^{(1)} = \underline{\sigma}_z$, $\underline{m}^{(2)} = \underline{\sigma}_x$, $\underline{m}^{(3)} = \underline{\sigma}_z \underline{\sigma}_x$, and , therefore, $\underline{m}^{(1)}\underline{m}^{(2)}$ is $\underline{m}^{(3)}$ itself and no problems appear.

---

* The case of N=2 is the family of Dirac matrices discussed in text books of Quantum Mechanics; see,*e.g.*, reference [32].

If we consider another basis $\underline{\tilde{m}}^{(\alpha)}$ with matrices $\underline{\tilde{m}}^{(0)} = 2^{-N/2}\underline{m}^{(0)} = 2^{-N/2}\underline{1}$ and $\underline{\tilde{m}}^{(\alpha)} = 2^{-N/2}\, i\underline{m}^{(\alpha)}$ for $\alpha \neq 0$, and define $\underline{\tilde{M}} = \sum_{\alpha=0}^{2^{2N}-1} c_\alpha \underline{\tilde{m}}^{(\alpha)}$ with $c_\alpha$ real and $\sum_{\alpha=0}^{2^{2N}-1} c_\alpha^2 = 1$, we have

$$\underline{\tilde{M}}^\dagger \underline{\tilde{M}} = 2^{-N}\left[\underline{1} + \sum_{\alpha\neq\beta;\alpha,\beta=1}^{2^{2N}-1} c_\alpha c_\beta \left(\underline{m}^{(\alpha)}\underline{m}^{(\beta)} + \underline{m}^{(\beta)}\underline{m}^{(\alpha)}\right)\right]. \tag{3.1}$$

Since all member $\underline{m}^{(\alpha)}$ and $\underline{m}^{(\beta)}$ (with $0 \neq \alpha \neq \beta \neq 0$) cannot anticommute, $\underline{\tilde{M}}$ cannot become $2^{-N}\underline{1}$, and hence perfect SQT with state $\left|\underline{\tilde{M}}\right\rangle$ is not possible.

Since $\underline{e}^{(\alpha)}$ and $\underline{\tilde{M}}$ both give orthonormal bases with basis states

$$\left|e^{(\alpha)}\right\rangle_{\{A\}\{B\}} = \sum_{j,k=0}^{2^N-1} e_{jk}^{(\alpha)} \left|j\right\rangle_{\{A\}} \left|k\right\rangle_{\{B\}}, \quad \left|\tilde{M}\right\rangle_{\{A\}\{B\}} = \sum_{j,k=0}^{2^N-1} \tilde{M}_{jk} \left|j\right\rangle_{\{A\}} \left|k\right\rangle_{\{B\}}, \tag{3.2}$$

they can be connected by a unitary transformation and we can write

$$\left|e^{(\alpha)}\right\rangle = \hat{V}\left|\tilde{M}\right\rangle, \quad \hat{V} \equiv \sum_{\alpha=0}^{2^{2N}-1} \left|e^{(\alpha)}\right\rangle\left\langle\tilde{M}\right|, \tag{3.3}$$

where operator $\hat{V}$ define a unitary matrix $\underline{V}$ with elements $V_{jkj'k'}$ with $(j,k,j',k')$ $= 0,1,2,3,\ldots..,2^{2N}-1$ and the elements satisfy

$$e_{jk}^{(\alpha)} = \sum_{j',k'=0}^{2^{2N}-1} V_{jkj'k'} \tilde{M}_{j'k'}. \tag{3.4}$$

If now $\underline{E} = \sum_\alpha c_\alpha \underline{e}^{(\alpha)}$, we have $E_{jk} = \sum_\alpha c_\alpha \sum_{j'k'} V_{jkj'k'} M_{j'k'}$, and therefore,

$$\underline{E}^\dagger \underline{E} = \underline{\tilde{M}}^\dagger \underline{\tilde{M}}, \tag{3.5}$$

which is proved easily using unitarity of *V*. This proves that no matrix $\underline{E} = \sum_\alpha c_\alpha \underline{e}^{(\alpha)}$ with $c_\alpha$'s having a global phase can satisfy $\underline{E}^\dagger \underline{E} = 2^{-N}\underline{1}$ and no associated state $\left|E\right\rangle$ can lead to perfect SQT. This makes clear non-existence of magic basis for N > 1.

It should also be noted that the Rigolins's magic states [20] for entangled 4 qubit states do not possess the magic in the sense of Hill & Wootters. This can be shown easily by seeing that the entangled state $|E\rangle = (1/\sqrt{2})(|g_1\rangle + |g_2\rangle)$ $= (1/\sqrt{2})(|0000\rangle + |1111\rangle)$ does not give $\underline{E}$ with $\underline{E}^{\dagger}\underline{E} = 2^{-N}\underline{1}$. Here $|g_1\rangle$ and $|g_2\rangle$ are Rigolin's magic states.

**4. Existence of Magic Partial-Bases for Entangled 2N Qubit States with N>1**

Argument of last section show that if we select any r mutually anticommuting matrices $\underline{M}_l\ (l = 1,2,3,....,r)$ from the family $\underline{m}^{(\alpha)}\ (\alpha = 0,1,2,....,2^{2N}-1)$, the family of unit matrix $\underline{1}$ and matrices $i\underline{M}_l$ ($l$ = 1,2,3,....,r) give a magic partial basis similar to those described by Prakash and Maurya recently [4], because if

$$\underline{\tilde{M}} = \sum_{l=0}^{r} c_l \underline{\tilde{m}}_l\ ,\ \underline{\tilde{m}}_0 = 2^{-N/2}\underline{1},\ \underline{\tilde{m}}_l = 2^{-N/2}\, i\underline{M}_l\ \ (l = 1,2,...r), \tag{4.1}$$

all $c_l$'s have the same global phase $\theta$, *i.e*, $c_l = |c_l| e^{i\theta}$ and $\sum_{l=0}^{r} |c_l|^2 = 1$ ,

$$\underline{\tilde{M}}^{\dagger}\underline{\tilde{M}} = \sum_{l,l'=0}^{r} |c_l||c_{l'}| \underline{\tilde{m}}_l^{\dagger} \underline{\tilde{m}}_{l'}$$

$$= 2^{-N}\left[\sum_{l=0}^{r} |c_l|^2 \underline{1} + \sum_{(l\neq l'),l,l'=0}^{r} |c_l||c_{l'}| \left(M_l M_{l'} + M_l M_l\right)\right] = 2^{-N}\underline{1} \tag{4.2}$$

as matrices $\underline{M}_l$ and $\underline{M}_{l'}$ are anticommuting for $0 \neq l \neq l' \neq 0$.

For N=2 case, *i.e.*, for entangled 4-qubit states, the matrices $\underline{m}^{(\alpha)}$ $(\alpha = 0,1,2,....15)$ may be written as

$$\underline{m}^{(\alpha)} = (\sigma_1^{(\alpha_1)} \otimes \sigma_2^{(\alpha_2)})\ , \tag{4.3}$$

where the decimal number $\alpha$ has quaternary representation $\alpha_1\alpha_2$, *i.e.*, $\alpha = 4\alpha_1 + \alpha_2$ with $\alpha_{1,2}$ taking values 0,1,2,3, we can then define matrices $\underline{m}^{(\alpha)}$ which obey $Tr(\underline{\tilde{m}}^{(\alpha)}\underline{\tilde{m}}^{(\beta)}) = \delta_{\alpha\beta}$ and therefore the corresponding orthonormal states are given by matrices,

$$\underline{\tilde{m}}^{(1,1)} = \frac{1}{2}\begin{pmatrix} I & 0 \\ 0 & I \end{pmatrix} = \underline{I}\ ,\ \left(\underline{\tilde{m}}^{(1,2)}, \underline{\tilde{m}}^{(1,3)}, \underline{\tilde{m}}^{(1,4)}\right) = \frac{1}{2}\begin{pmatrix} \sigma & 0 \\ 0 & \sigma \end{pmatrix} = \left(\underline{A}_1, \underline{A}_2, \underline{A}_3\right),$$

$$\underline{\tilde{m}}^{(2,1)} = \frac{1}{2}\begin{pmatrix} I & 0 \\ 0 & -I \end{pmatrix} = \underline{F},\ \left(\underline{\tilde{m}}^{(2,2)}, \underline{\tilde{m}}^{(2,3)}, \underline{\tilde{m}}^{(2,4)}\right) = \frac{1}{2}\begin{pmatrix} \sigma & 0 \\ 0 & -\sigma \end{pmatrix} = (\underline{B}_1, \underline{B}_2, \underline{B}_3),$$

$$\underline{\tilde{m}}^{(3,1)} = \frac{1}{2}\begin{pmatrix} 0 & I \\ I & 0 \end{pmatrix} = \underline{G},\ \left(\underline{\tilde{m}}^{(3,2)}, \underline{\tilde{m}}^{(3,3)}, \underline{\tilde{m}}^{(3,4)}\right) = \frac{1}{2}\begin{pmatrix} 0 & \sigma \\ \sigma & 0 \end{pmatrix} = (\underline{C}_1, \underline{C}_2, \underline{C}_3),$$

$$\underline{\tilde{m}}^{(4,1)} = \frac{1}{2}\begin{pmatrix} 0 & -iI \\ iI & 0 \end{pmatrix} = \underline{H},\ \left(\underline{\tilde{m}}^{(4,2)}, \underline{\tilde{m}}^{(4,3)}, \underline{\tilde{m}}^{(4,4)}\right) = \frac{1}{2}\begin{pmatrix} 0 & -i\sigma \\ i\sigma & 0 \end{pmatrix} = (\underline{D}_1, \underline{D}_2, \underline{D}_3).$$

Explicitly, the sixteen orthonormal states are

$$\underline{I} \Rightarrow \frac{1}{2}\left[|0000\rangle + |0101\rangle + |1010\rangle + |1111\rangle\right] = |I\rangle,$$

$$\underline{F} \Rightarrow \frac{1}{2}\left[|0000\rangle + |0101\rangle - |1010\rangle - |1111\rangle\right] = |F\rangle,$$

$$\underline{G} \Rightarrow \frac{1}{2}\left[|0010\rangle + |0111\rangle + |1000\rangle + |1101\rangle\right] = |G\rangle,$$

$$\underline{H} \Rightarrow -\frac{i}{2}\left[|0010\rangle + |0111\rangle - |1000\rangle - |1101\rangle\right] = |H\rangle,$$

$$\underline{A}_1 \Rightarrow \frac{1}{2}\left[|0001\rangle + |0100\rangle + |1011\rangle + |1110\rangle\right] = |A_1\rangle,$$

$$\underline{A}_2 \Rightarrow -\frac{i}{2}\left[|0001\rangle - |0100\rangle + |1011\rangle - |1110\rangle\right] = |A_2\rangle,$$

$$\underline{A}_3 \Rightarrow \frac{1}{2}\left[|0000\rangle - |0101\rangle + |1010\rangle - |1111\rangle\right] = |A_3\rangle,$$

$$\underline{B}_1 \Rightarrow \frac{1}{2}\left[|0001\rangle + |0100\rangle - |1011\rangle - |1110\rangle\right] = |B_1\rangle,$$

$$\underline{B}_2 \Rightarrow -\frac{\mathrm{i}}{2}\left[|0001\rangle - |0100\rangle - |1011\rangle + |1110\rangle\right] = |B_2\rangle,$$

$$\underline{B}_3 \Rightarrow \frac{1}{2}\left[|0000\rangle - |0101\rangle - |1010\rangle + |1111\rangle\right] = |B_3\rangle,$$

$$\underline{C}_1 \Rightarrow \frac{1}{2}\left[|0011\rangle + |0110\rangle + |1001\rangle + |1100\rangle\right] = |C_1\rangle,$$

$$\underline{C}_2 \Rightarrow -\frac{\mathrm{i}}{2}\left[|0011\rangle - |0110\rangle + |1001\rangle - |1100\rangle\right] = |C_2\rangle,$$

$$\underline{C}_3 \Rightarrow \frac{1}{2}\left[|0010\rangle - |0111\rangle + |1000\rangle - |1101\rangle\right] = |C_3\rangle,$$

$$\underline{D}_1 \Rightarrow -\frac{\mathrm{i}}{2}\left[|0111\rangle + |0110\rangle - |1001\rangle - |1100\rangle\right] = |D_1\rangle,$$

$$\underline{D}_2 \Rightarrow \frac{1}{2}\left[-|0011\rangle + |0110\rangle - |1001\rangle + |1100\rangle\right] = |D_2\rangle,$$

$\underline{D}_3 \Rightarrow -\frac{i}{2}\left[|0010\rangle - |0111\rangle - |1000\rangle + |1101\rangle\right] = \left|D_3\right\rangle$.

One can write multiplication table for these matrices and see that the sets of mutually anticommuting matrices with maximum possible elements are {$\underline{F}$, $\underline{G}$, $\underline{D}_1$, $\underline{D}_2$, $\underline{D}_3$}, {$\underline{G}$, $\underline{H}$, $\underline{B}_1$, $\underline{B}_2$, $\underline{B}_3$},{$\underline{H}$, $\underline{F}$, $\underline{C}_1$, $\underline{C}_2$, $\underline{C}_3$}, {$\underline{A}_1$, $\underline{A}_2$, $\underline{B}_3$, $\underline{C}_3$, $\underline{D}_3$}, {$\underline{A}_2$, $\underline{A}_3$, $\underline{B}_1$, $\underline{C}_2$, $\underline{D}_2$}, {$\underline{A}_3$, $\underline{A}_1$, $\underline{B}_2$, $\underline{C}_2$, $\underline{D}_2$}, {$\underline{F}$, $\underline{G}$, $\underline{H}$,} and {$\underline{A}_1$, $\underline{A}_2$, $\underline{A}_3$},the magic partial bases are the sets of states {$|I\rangle$, $i|F\rangle$, $i|G\rangle$, $i|D_1\rangle$, $i|D_2\rangle$, $i|D_3\rangle$}, {$|I\rangle$, $i|G\rangle$, $i|H\rangle$, $i|B_1\rangle$, $i|B_2\rangle$, $i|B_3\rangle$}, {$|I\rangle$, $i|H\rangle$, $i|F\rangle$, $i|C_1\rangle$, $i|C_2\rangle$, $i|C_3\rangle$}, {$|I\rangle$, $i|A_1\rangle$, $i|A_2\rangle$, $i|B_3\rangle$, $i|C_3\rangle$, $i|D_3\rangle$}, {$|I\rangle$, $i|A_2\rangle$, $i|A_3\rangle$, $i|B_1\rangle$, $i|C_1\rangle$, $i|D_1\rangle$}, {$|I\rangle$, $i|A_3\rangle$, $i|A_1\rangle$, $i|B_2\rangle$, $i|C_2\rangle$, $i|D_2\rangle$}, {$|I\rangle$, $i|F\rangle$, $i|G\rangle$, $i|H\rangle$} and {$|I\rangle$, $i|A_1\rangle$, $i|A_2\rangle$, $i|A_3\rangle$}. The maximum dimension of magic partial-bases is 6 and hence magic semi-bases (with dimension $\frac{1}{2}\times 16 = 8$) [9] do not exist. However the magic quarter-bases which have $\frac{1}{4}\times 16 = 4$ basis states do exist and these having no common states except $|I\rangle$ are {$|I\rangle$, $i|F\rangle$, $i|G\rangle$, $i|H\rangle$}, {$|I\rangle$, $i|A_1\rangle$, $i|A_2\rangle$, $i|A_3\rangle$}, {$|I\rangle$, $i|B_1\rangle$, $i|B_2\rangle$, $i|B_3\rangle$}, {$|I\rangle$, $i|C_1\rangle$, $i|C_2\rangle$, $i|C_3\rangle$} and {$|I\rangle$, $i|D_1\rangle$, $i|D_2\rangle$, $i|D_3\rangle$}.

## 5. Conclusion

In section 2, we gave a simple and precise protocol for teleporting faithfully an arbitrary N-qubit state from Alice to Bob using 2N-qubit entangled states. We noted that the entangled state $|E\rangle$ and the Bell states $\left|B^{(\alpha)}\right\rangle$ define matrices $\underline{E}$ and $\underline{B}^{(\alpha)}$, which satisfy $\underline{E}^{\dagger}\,\underline{E} = \underline{B}^{(\alpha)\dagger}\,\underline{B}^{(\alpha)} = 2^{-N}\underline{1}$. This suggests that $\underline{E}^{\dagger}\,\underline{E} = 2^{-N}\underline{1}$ may be taken as criterion for perfect entanglement for entangled 2N-qubit states (see also [14]). We also saw how concepts of character matrix and transformation, judgment and kernel operators come in a natural way in our simple theory. In section 3, we proved rigorously non-existence of Hill-Wootters type magic basis for entangled 2N-qubit states for N > 1 and in section 4, we proved existence of magic partial bases. We also found explicitly these magic partial bases for N = 2.

## Acknowledgement

We are thankful to Prof. N. Chandra and Prof. R. Prakash for their interest and critical comments and to Dr. R. Kumar, Dr. P. Kumar, Dr. D.K. Mishra, N. Shukla, A. K. Yadav, A. K. Maurya, and M. K. Mishra for helpful and stimulating discussions. One of the authors would like to thanks University Grant Commission also for financial support.